\documentclass[aps,preprint,twocolumn,superscriptaddress,10pt]{revtex4}
\usepackage{amsfonts}
\usepackage{amsmath}
\usepackage{amssymb}
\usepackage{graphicx}
\usepackage{dcolumn}
\usepackage{mciteplus}

\begin{document}

\title[Spin excitations in K$_{2}$Fe$_{4+x}$Se$_{5}$: linear response $ $
approach]{Spin excitations in K$_{2}$Fe$_{4+x}$Se$_{5}$: linear response
approach}
\author{Liqin Ke}
\affiliation{Ames Laboratory USDOE, Ames, IA 50011}
\author{Mark van Schilfgaarde}
\affiliation{Department of Physics, King's College London, Strand, London WC2R 2LS}
\author{Vladimir Antropov}
\affiliation{Ames Laboratory USDOE, Ames, IA 50011}
\pacs{PACS number}

\begin{abstract}
Using \emph{ab initio} linear response techniques we calculate spin wave
spectra in K$_{2}$Fe$_{4+x}$Se$_{5}$, and find it to be in excellent
agreement with a recent experiment. The spectrum can be alternatively
described rather well by localized spin Hamiltonian restricted to first and
second nearest neighbor couplings. We confirm that exchange coupling between
nearest neighbor Fe magnetic moments is strongly anisotropic, and show
directly that in the ideal system this anisotropy has itinerant nature which
can be imitated by introducing higher order terms in effective localized
spin Hamiltonian (biquadratic coupling). In the real system, structural
relaxation provides an additional source of the exchange anisotropy of
approximately the same magnitude. The dependence of spin wave spectra on
filling of Fe vacancy sites is also discussed.
\end{abstract}

\eid{identifier}
\date{\today }
\maketitle

Because there seems to be a close, if poorly understood, connection between
magnetic excitations and superconductivity in the recently discovered families
of Fe-based superconductors, a great deal of attention has been paid to the
elementary magnetic excitations in these systems\cite {REVJohnson,RevNeut}.
While the number of materials with superconducting properties is rapidly
growing, they all share in common a phase diagram with an antiferromagnetic
region immediately adjacent to the superconducting phase.  AFM order in most
parent compounds are characterized by rather steep spin waves (antiferromagnons)
and relatively low N{\'{e}}el temperature ($\sim$140\,K)
\cite{REVJohnson,RevNeut}.  On the other hand, iron selenides
{K$_{2}$Fe$_{4+x}$Se$_{5}$} discovered very recently, which we will call here a
``245'' system, is a particularly interesting case, because while its
superconducting state appears to be similar in many respects to some of the
other families, has a much higher N{\'{e}}el temperature ($T_{N}{\sim }560K$)
and very large magnetic moment (${\sim}$3$\mu_{B}$) \cite{BAO1,BAO2,BAO3}.  The
first neutron experiments, performed only recently~\cite{245SW}, show collective
spin excitations somewhat similar to excitations in the Fe-pnictides such as
CaFe$_{2}$As$ _{2} $.  A highly debated topic of discussion is the anisotropy of
the exchange coupling.  In Ref.~\cite{245SW} the authors fit a set of
anisotropic Heisenberg model parameters to their data, and argued that at least
three effective nearest neighbor (NN) exchange parameters are needed to fit
observed spectra in a satisfactory manner.  In another recent work \cite {HUBIQ}
this exchange anisotropy has been fully attributed to biquadratic exchange,
supporting the model introduced in Ref.~\cite{245SW}.  However, neutron linear
response experiments do not contain sufficient information to establish whether
or not higher order terms of localized spin Hamiltonian or other interactions
are responsible for this anisotropy.  Information from the linear regime cannot
unambiguously distinguish the dependence of linear response behavior on
environment and terms originating from higher order.  The unique answer can be
obtained using non-collinear band structure calculations together with a linear
response theory \cite{LR}.

Here we analyze spin excitations in the 245 system using non-collinear density
functional theory in tandem with a linear response approach to determine the
spin wave (SW) spectra of both the parent compound, {K$_{2}$Fe$_{4}$Se$_{5}$},
and its modification by addition of Fe on the `vacancy' sites as described
below.  We accurately reproduce observed features in SW spectra of the parent
compound from this first principles approach, without recourse to empirical or
adjustable parameters.  Thus we argue that the Heisenberg model extracted from
this theory provides a good description of the static transverse susceptibility,
and the exchange parameters $J_{ij}$ derived \emph{ab initio} from it, which
have a physical meaning as the second variation in total energy with respect to
spin rotations, are accurately given by the theory.  We can also unambiguously
address what happens when exchange parameters, which can be computed to whatever
range is desired, are truncated to a short range, e.g. just first and second
neighbors ($J_{1}${--}$J_{2}$ model).

For computational convenience we adopt a multiple-scattering approach in the
long wave approximation~\cite{JLWA}, which is reasonable provided the local
Fe moment is sufficiently large.  This is the case in {K$_{2}$Fe$_{4+x}$Se$_{5}$}, where the Fe local moment is both measured to be $\sim $3$\mu _{B}$,
and predicted to be so in density-functional theory.

\begin{figure*}[tbph]
\includegraphics[scale=0.3]{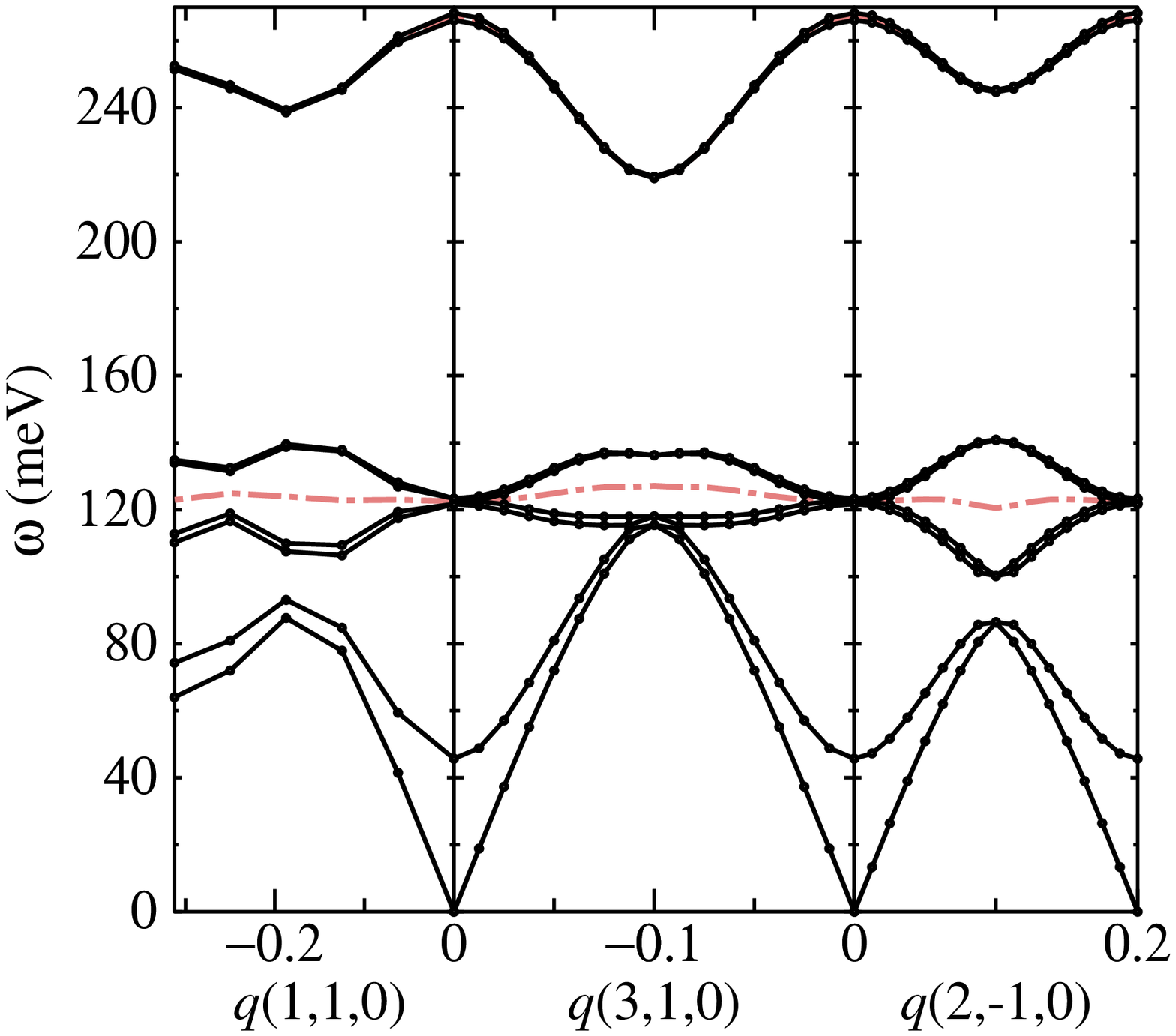} \quad 
\includegraphics[scale=0.55]{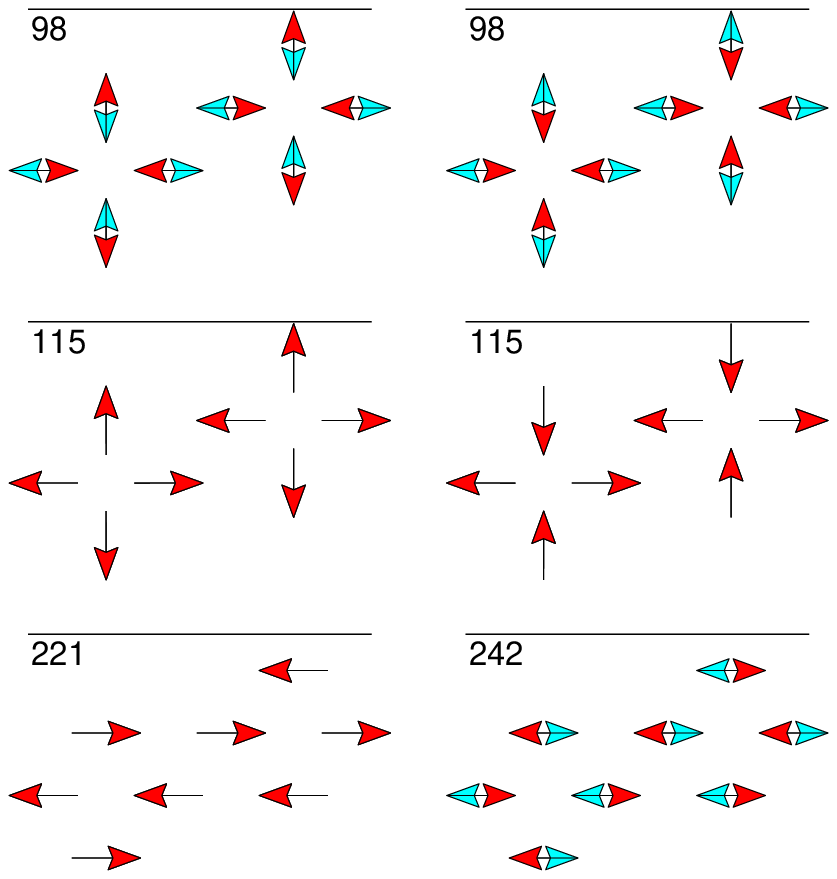} \quad 
\includegraphics[scale=0.3]{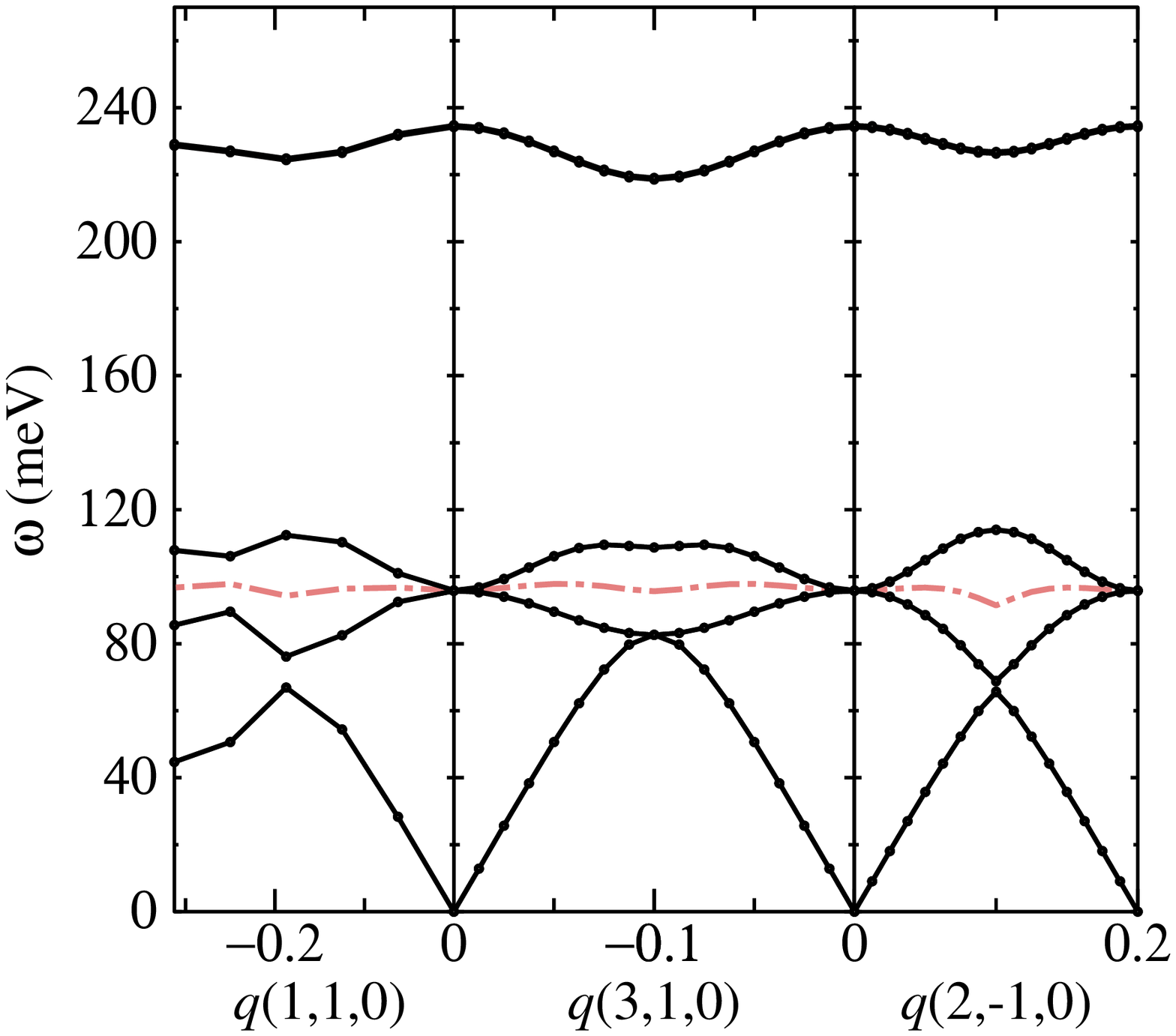}
\caption{Spin-wave spectra in K$_{2}$Fe$_{4}$Se$_{5}$ calculated from the
static susceptibility in the LDA.  Panel $(a)$ shows $\protect\omega (q)$,
with the right frame for $q$ along the $\Gamma {-}G_{1}$ line.  It
corresponds to the experimental spectra shown in Ref.~\protect\cite{245SW}
(bottom panels of Fig. 3), whereas the left frame corresponds to the top
panel of that Figure.  The middle frame shows spectra along the $\Gamma {-}
(G_{1}{+}G_{2})$ line.  Panel $(b)$ shows eigenvectors of the six optical
modes at $\Gamma $, and their corresponding frequencies (meV).  The
perspective is from the $z$ axis, so each point contains two Fe atoms.
Arrows corresponding to spin rotations of each of the 16 Fe atoms.  In some
panels only 8 arrows are visible; this occurs when the rotations in the
upper and lower planes are in phase.  In other panels all 16 arrows are
visible; for these modes the lower and upper planes are 180$^{\circ }$ out
of phase.  The right frame shows how the SW get modified when the exchange
interactions are restricted to $J_{1}$ and $J_{2}$ only.}
\label{fig:swparent}
\end{figure*}

The parent compound {K$_{2}$Fe$_{4}$Se$_{5}$} has a vacancy on the Fe sublattice
for each formula unit, which if filled would have a composition
K$_{2}$Fe$_{5}$Se$_{5}$ and be structurally similar to the other Fe
superconductors.  The ordered magnetic phase of K$_{2}$Fe$_{4}$Se $_{5}$ is
unusual: blocks of four Fe atoms are coupled ferromagnetically in a square; the
squares are arranged antiferromagnetically in a N{\'{e}}el like configuration,
which we denote the ``block N{\'{e}}el'' magnetic structure.  In what manner the
vacancy Fe sites are filled in superconducting material is a crucial issue,
because the parent compound is predicted in density-functional theory to be an
insulator with a bandgap of $\sim $0.4~eV \cite{CAO}.

\begin{table}[tbp]
\caption{Bond lengths (\AA) of K$_{2}$Fe$_{4}$Se$_{5}$ in the block
N{\'{e}} structures calculated from PBE and LDA functionals, bandgap (eV)
and magnetic moment $M$ ($\protect\mu_{B}$). ``LDA@PBE'' refers to
calculations using the Barth-Hedin (local) functional, but calculated at
the PBE minimum-energy geometry.  The shorter (longer) of the Fe-Fe bond
length is the intrablock (interblock) distance.  Mean PBE Fe-Fe and Fe-Se
bond lengths are close to the experimental values.  ``25\% Fe'' refers to
an LDA@PBE calculation with 25\% doping of the Fe vacancy sites, as
discussed in the text.}
\label{tab:tab1}
\begin{tabular}{|c|c|c|c|c}
\  & $d$({Fe-Fe}) & $d$({Fe-Se}) & $E_{g}$ & $M$ \\ 
\hline
Expt & 2.768          & 2.441        &      & 3.3   \\ \hline
PBE  & 2.657--2.901   & 2.406--2.509 & 0.59 & 3.02  \\ \hline
LDA  & 2.649--2.872   & 2.354--2.454 & 0.26 & 2.75  \\ \hline
LDA@PBE &             &              & 0.44 & 2.90  \\ \hline 
25\% Fe & 2.604--2.814& 2.399--2.518 & 0.01 & 2.3--2.9 \\ \hline
\end{tabular}
\end{table}

We first consider magnetic excitations of the parent compound, with
reciprocal lattice vectors $G_{1}$=(0.4,${-}$0.2,0), $G_{2}$=(0.2,0.4,0),
and $G_{3}$=(0,0,$a/c$), in units of $2\pi /a$.  This cell contains four
formula units of {K$_{2}$Fe$_{4}$Se$_{5}$} with two units per plane.  We used
experimental lattice constants from Ref. \cite{KFESESTRUC} with $a$=3.914\AA 
, and $c/a$=3.587.  Because the LDA is known to underestimate Fe-chalcogen
and Fe-pnictogen bond lengths in these families of superconductors, we use
the PBE functional to relax the crystal structure in the block N{\'{e}}el
magnetic state.  The resulting relaxed structure, is seen in Table~\ref
{tab:tab1} to agree well with experimental data.  On the other hand, the PBE
functional tends to overestimate the magnetic moments and magnetic
stabilization energy.  Thus to calculate magnetic excitations, we use the
LDA, but adopt the crystal structure predicted by the PBE functional.

The full susceptibility $\chi^{-1}(\mathbf{q},\omega )$ is prohibitively
expensive to calculate for such a large system.  Thus we adopt an approach
which relies on the atomic sphere approximation (ASA) following Ref.~\cite
{vanschilfgaarde.jap1999,*antropov.jap2006}.  To check ASA approximation to DFT, we carefully checked the
energy bands, density of states, and magnetic moment in ASA against LDA
results and found satisfactory agreement for all cases described here; for
example the ASA bandgap and magnetic moment of the parent compound were
found to be 0.40~eV and 2.85$\mu _{B}$, close to the LDA@PBE result of Table~
\ref{tab:tab1}.

Pair exchange parameters $J_{ij}$ were obtained from a Fourier transform of
$J(\mathbf{q})$; the latter was calculated on fine a $q$-mesh of {16$\times
$16$\times $4} divisions in order to reliably extract $J_{ij}$ to very
distant neighbors.  Resulting spin wave spectra for the parent compound
K$_{2} $Fe$_{4}$Se$_{5}$ are shown in Fig.~\ref{fig:swparent}.  Focusing
first on the $\Gamma $ point, six optical modes are seen: two degenerate
pairs clustered around 100~meV, and two others near 230 meV. (We defer
discussion of the mode near 40~meV for the moment.) The middle panel of
Fig.~\ref {fig:swparent} depicts their eigenvectors.  The lowest pair and
second pair are essentially the same in a given plane, but top and bottom
planes rotate against each other for the low-energy pair, and in phase for
the high energy pair.  The first high-energy mode consists uniform
rotations spins against parallel sheets on the $xz$ axis; the second is
similar but upper and lower planes are out of phase.

Left and right frames can be compared directly with neutron data, top and bottom
panels of Fig. 3 of Ref.~\cite{245SW}.  All modes match experimental data very
well, when broadening in the experimental spectrum is taken into account.  The
acoustic mode, with linear dispersion for small $q$ of Fig.~\ref {fig:swparent}
has a maximum of 86 meV at $q{=}(1/2)G_{1}$, slightly larger than the measured
value in Ref.~\cite{245SW}.  The four low-energy optical modes are all
``breathing'' modes: spins rotate radially away from the center of their square.
These four separate modes cannot be resolved in experiment; instead the measured
data reflects a superposition of these modes.  It is nearly dispersionless along
the $\Gamma { -}G_{1}$ line, with $\omega {\approx }110$~meV.  For comparison to
experiment four optical modes were averaged and shown as a dashed line in
Fig.~\ref {fig:swparent}, and the result is seen to be the same as the neutron
data, though at slightly higher energy.  Finally the two high-energy optical
modes show a weak $q$ dependence, with $\omega {\approx }240$~meV.  Essentially
identical behavior is observed in the neutron data along the (110) line, though
$\omega $ is slightly smaller.  The N{\'{e}}el temperature, $T_{N}$, estimated
within the RPA, is about 12\% smaller than the observed value.  This is in fact
very good agreement with experiment, since the RPA is known to underestimate
critical temperatures for a given Hamiltonian, by roughly 10\%.

Finally, consider the mode appearing near 40~meV.  If the two Fe planes
were identical and uncoupled this mode would exactly coincide with the
acoustic mode.  Thus it is essentially the acoustic mode with
$q_{z}{=}1/2\cdot {2\pi } /c$, zone-folded to the $\Gamma $ point, because
the actual crystal contains two planes of Fe.  The splitting of this and
the true acoustic mode are a result of coupling between the two planes
through exchange parameters along $J^{z}$ coupling different planes.  This
shows that $J^{z}$ is small, but not so small that planes are decoupled.


\begin{table}[tbp]
\caption{Exchange energies $J_{k}$, in meV, for the Heisenberg hamiltonian
$H{=}\sum J_{ij}\,\mathbf{S}_{i}\cdot\mathbf{S}_{j}$, of the parent
compound K$_{2}$Fe$_{4}$Se$_{5}$.  Here $k$ denotes the neighbor
index. $J_{k}$ and $J_{k}^{\prime}$ refer to couplings between pairs of
like spin and opposite spin, respectively.  Also shown is the N{\'{e}}el
temperature, estimated from the RPA \protect\cite{TYABLIKOVMANYATOMS}.}
\label{tab:tab2}
\begin{tabular}{|c|r|c|c|c|}
& \multicolumn{2}{c|}{Relaxed geometry} & \multicolumn{2}{c|}{Ideal geometry}\\
\ $k$ & $J_{k}$ & $J^{\prime}_{k}$ & $J_{k}$ & $J^{\prime}_{k}$ \\
\hline
  1   & -9.7  & 27.3                  & -5.7  & 25.7     \\ \hline
  2   & 10.2  &  8.5                  & 17.0  &  6.7                 \\ \hline
  3   &  0.3  & -0.3                  &  0.1  & -0.1                 \\ \hline
  4   & -1.9  &  0.1\rlap{$\pm1.0$}   & -2.2  &  0.5\rlap{$\pm0.5$}  \\ \hline
  5   &  0.1  &  0.5                  & -0.3  &  0.5                 \\ \hline
  6   & -0.1  &  0.1\rlap{$\pm0.1$}   &       &  0.1\rlap{$\pm0.1$}  \\ \hline
  7   & -0.1  &  1.3                  &       &               \\ \hline
$T_N$ & \multicolumn{2}{c|}{494~K} & \multicolumn{2}{c|}{\hbox{\quad}(not stable)\hbox{\qquad}} \\
\hline
\end{tabular}
\end{table}

A great deal of speculation has arisen in the community about the size and
range of parameters $J$.  Much effort has been expended extracting
effective exchange parameters $J_{ij}$ by fitting a Heisenberg model with a
few (2 or 3) neighbors to the observed data, as was done in
Refs.~\cite{245SW} and \cite{HUBIQ}.  Since, as we have shown, SW can well
described by the Heisenberg model derived \emph{ab initio}, these results
enable us to unambiguously address questions about the range and
environment dependence of $J_{ij}$.  In particular, it is debated whether a
$J_{1}$--$J_{2}$ model is sufficient or more distant neighbors are
required.  Table~\ref{tab:tab2} shows $J_{ij}$ for the first few neighbors,
distinguishing intra-block and interblock couplings.  As widely thought,
the first two neighbors give by far the largest contribution to $J$.  To
quantify this effect, we recalculated the SW spectrum with $J$ restricted
to only nearest and second neighbor.  This leaves four independent
parameters since interblock $J$ and intrablock $J^{\prime}$ are distinct
(Table~\ref{tab:tab2}).  The resulting SW spectra, shown in the right panel
of Fig.~\ref{fig:swparent}, look qualitatively similar to the full
calculation, except that some modes become degenerate, and low-energy modes
soften by about 25\%.  The predicted $T_N$ drops by a comparable amount,
from 494K to 399K.  Perhaps most importantly, the ``zone-folded'' acoustic
mode noted above becomes degenerate with the true acoustic mode, because
all interplane interactions are now excluded by construction.

Remarkably, corrections to the $J_{1}$--$J_{2}$ approximation receive
almost \emph{no} contribution from $J_{3}$; rather, they originate from
$J_{4}$--$J_{7}$ (especially $J_{7}^{\prime}$).  The magnetic structure is
stabilized by strong AFM first and second neighbor interblock coupling, and
weakly so by nearest intrablock exchange.  Frustration is present, since
the AFM second neighbor intrablock coupling acts strongly to destabilize
the magnetic order, but it is overcome because there are half as many such
pairs as their interblock counterparts.  The Table also shows what $J_{ij}$
would obtain if the structure were constrained to an ideal geometry.
Relaxation causes a small deformation of the Fe atoms in squares; there is
also some dispersion in the Fe-Se bond length around the average value
(Table~\ref {tab:tab1}).  This structural relaxation is critically
important for magnetic stability.  As Table~\ref{tab:tab2} shows, the
stability of the (intrablock) square is strongly enhanced by relaxation:
the NN coupling is increased and frustration in the 2nd NN is strongly
reduced.  Without relaxation some of SW frequencies become complex,
indicating that the collinear state is not stable.

\begin{figure}[hptb]
\includegraphics[scale=0.4]{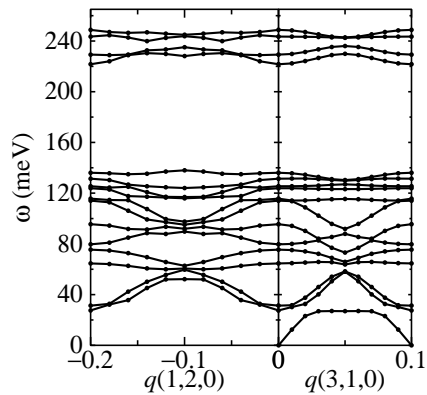}
\caption{Spin-wave spectra for {K$_{2}$Fe$_{4.25}$Se$_{5}$} as described in
the text.  Actual cell is K$_{16}$Fe$_{34}$Se$_{40}$.}
\label{fig:swfe25}
\end{figure}

As the filling of Fe vacancy sites very likely play an essential role in
mediating superconductivity, how elementary excitations are modified when
these sites are partially occupied is a crucial question.  We consider
first 25\% occupation of Fe vacancies.  To retain an overall zero-spin
configuration it is necessary to double the size of the four formula unit
cell: we construct a supercell with $G_{1}$=(0.3,0.1,0) and
$G_{2}$=(${-}$0.1,0.3,0), and populate two of the eight vacancy sites with
a pair of Fe$^{\uparrow}$ and Fe$^{\downarrow}$ atoms in such a way as to
preserve a two-fold rotational symmetry.  The lattice was relaxed using the
PBE functional, with the resulting equilibrium bond lengths shown in
Table~\ref {tab:tab1}.  SW's are shown in Fig.~\ref{fig:swfe25}.  The
acoustic mode (now zone folded) is largely unchanged and uncoupled from the
other modes: there is a linear dispersion at small $q$ and a maximum near
80~meV for, e.g. $\mathbf{q}$=(0.3,0.1,0) and $\mathbf{q}$=(0.1,$-$0.2,0).
The (zone-folded) low-energy optical modes now show some new features.  In
addition to the modes in the 80-120 meV range, new and largely
dispersionless modes appear around 60 meV.  The eigenfunctions of these
modes are very complex, with strong admixtures of both vacancy sites and
host sites.  The high-frequency modes are almost unchanged as ``vacancy''
Fe atoms participate very little in them.

At 25\% filling the magnetic structure is stable; indeed $T_{N}$ for
{K$_{2} $Fe$_{4}$Se$_{5}$}, predicted to be 494\,K in the Tyablikov
approximation, is only slightly reduced, to 485K for
{K$_{2}$Fe$_{4.25}$Se$_{5}$}.  We also studied a case with 50\% filling.  A
collinear magnetic state was stabilized; however the linear response
calculation of the SW spectra revealed imaginary frequencies, indicating
that the collinear state is not stable.

The exchange anisotropy, $J_{k}{-}J_{k}^{\prime }$, is unusually large in
this material.  It is natural to associate $J_{k}{-}J_{k}^{\prime }$ with
biquadratic coupling, as was done in Refs.\cite{JIJUS,*BIQNATURE}.  However,
lattice relaxations are very important in this case; compare the ideal to
relaxed $J$ in Table~\ref{tab:tab2}.  Our calculations predict that the
intrablock NN Fe-Fe bond length is calculated to be smaller than the
intrablock one by $\sim$8\% (Table I); and perhaps more important, there is
a significant dispersion in the Fe-Se bond lengths.  The strictly
electronic part of biquadratic contribution is still present and our
calculations for the ideal 245 (no relaxation) show that biquadratic
coupling is about 30\% of $J_{1}$.  This result was obtained by set of
calculations of non-collinear configurations that enable us to go beyond
linear response.  We fit the total energy of the different non-collinear
states to effective spin Hamiltonian with higher order terms and found a
large cos$^{2}\theta $ contribution from first NN, while contribution from
higher harmonics appeared to be smaller.  Also biquadratic coupling from
the second NN is smaller.  Thus the environment dependence of $J$
originates from two sources: purely electronic terms for a fixed lattice,
and an ``exchange-striction'' contribution, originating from
spin-configuration-dependent lattice relaxation.

These non-collinear calculations point to the microscopic origin of the
biquadratic interaction.  It appears that major contribution to this
'biquadratic' energy is produced by the change of the amplitude of magnetic
moments.  $M$ was found to vary by $\sim$0.3$\mu_{B}$, even though its
magnitude is large and thought to be well described by a rigid local-moment
picture.  The presence of unusually large interaction between longitudinal
and transverse degrees of freedom reflects the relatively large itinerant
component of magnetic interactions in this system.  It is notable that the
interaction between these kinds of fluctuations lies outside a linear
response description.  The LDA is a mean-field theory and does not
incorporate fluctuations directly.  It can nevertheless capture some aspect
of fluctuations through the imposition of constraints (spin orientations).

In conclusion, we demonstrated that \textit{ab-initio} linear response
method nearly perfectly describes the observed spin wave spectra in the
complicated 245 system. The experimental spectrum alternatively can be
described reasonably well by localized spin Hamiltonian with two nearest
neighbor couplings. We confirm the anisotropy of the exchange coupling
between NN Fe magnetic moments and established that this anisotropy is
associated with significant biquadratic coupling which appears in part
because of longitudinal fluctuations. Structural relaxation provides an
additional source of the exchange anisotropy of approximately the same
magnitude.

Work at the Ames Laboratory was supported by DOE Basic Energy Sciences,
Contract No. DE-AC02-07CH11358.

\bibliography{bbb}

\end{document}